# The Marginal Stability of Proteins: How the Jiggling and Wiggling of Atoms is Connected to Neutral Evolution

Osvaldo A. Martin[1] and Jorge A. Vila[1]

[1] IMASL-CONICET, Universidad Nacional de San Luis, Ejército de Los Andes 950, 5700 San Luis, Argentina.

Abstract

Here we propose that the upper bound marginal stability of proteins is a universal property that includes macro-molecular complexes and is not affected by molecular changes such as mutations and post-translational modifications. We theorize that its existence is a consequence of Afinsen's thermodynamic hypothesis rather than a result of an evolutionary process. This result enables us to conjecture that neutral evolution should also be, with respect to protein stability, a universal phenomenon.

We are interested in studying the marginal stability of proteins from a thermodynamics perspective and its connection to neutral evolution. Following Hormoz[1] we understand that protein marginal stability "…*refers to the thermodynamic stability and is equivalent to the size of the energy gap (or the energy difference) between the native state and the first excited (misfolded) state*…" That is, we are interested in biologically active proteins and complexes following a dynamic exchange with slightly higher-energy conformations retaining biological activity. Hence, we are not interested in the analysis of the protein stability understood as the free energy of denaturation. Having clarified this essential point, we wonder whether neutral evolution leads to marginally stable proteins or whether the latter are a necessary consequence of Anfinsen's thermodynamic hypothesis, that sets "…*the three-dimensional structure of a native protein in its normal physiological milieu ... is the one in which the Gibbs free energy of the whole system is lowest...*" [2] This is a central question that deserves to be clarified as the origin of protein evolvability is a long-lasting problem and a major challenge to the field of evolutionary biology and, despite the existence of a large body of theoretical investigations, there is not a conclusive and clear answer to this inquiry,[3-12] yet.

In a previous work[13] we have been able to demonstrate, based on the use of simple concepts from statistical thermodynamics and the Gershgorin theorem, the existence of an upper limit to the marginal stability of monomeric globular proteins, namely ≈7.4 Kcal/mol. We had also provided sound arguments that this upper limit to the marginal stability is a consequence of a quasi-equilibrium of forces that take place at the minimum of the protein global free energy. In addition, because there is no condition on the proteins other than being monomeric and globular the above result is valid for any fold-class, sequence or proteins size. Here, we largely expand this analysis to show that this upper-bound limit of ≈7.4 Kcal/mol is broadly supported by new studies in different sized molecules/complexes. Therefore, this upper-bound limit seems to be a universal property of proteins and macromolecular complexes and, consequently, it should not be affected by molecular changes, such as mutations and/or Post-Translational Modifications (PTMs). Let us discuss 3 studies that are supportive of these conjectures.

1. A recent study of the ribosome native state shows that the largest energy difference, between the lowest and highest point of the free energy landscape, is ≈3.80 ± 0.65 Kcal/mol.[14] This means that proteins with a molecular weight of about $2.5 \times 10^4$ Da, such as α-Chymotripsin,[15] and the ribosome, a molecular complex of 79 proteins and 4 RNAs, with a molecular weight of about $3.2 \times 10^6$ Da both have similar marginal stabilities. Certainly, the results from these studies can be rationalized by looking at the definition of the upper bound free energy change derived for the native-state of monomeric globular proteins:[13]

$$\Delta G \leq \mathcal{L}im_{MW \to \infty} RT \ln MW \qquad (1)$$

where *MW* stands for the molecular weight, *R* the gas constant and *T* the temperature. Because of the logarithm in (1), ΔG is robust upon changes in the value of *MW*. Indeed, at room temperature an increase of *MW* by one order of magnitude is going to result into a ΔG increase of ≈1.4 Kcal/mol. This quantity is similar to the average strength of protein's hydrogen bonds in solution, i.e. ≈1.5 Kcal/mol.[16] Moreover, even if we do not use *MW* in (1), and instead we estimate the free-energy change as some unknown function of molecular features like number of hydrogen bonds, residue-residue contacts, etc., the robustness will still hold because, for example, the number of intra main-chain hydrogen bonds roughly scales as *2N*, with *N* being the number of amino-acids, while the number of contacts of pairwise and multi-body interactions also scales linearly with *N*. This implies that the stability upper bound should be roughly similar, i.e., in the order of a few hydrogen bonds of difference, for at least most of the functional bio-molecules and bio-molecular complexes.

2. Analysis of protein stability changes in response to protein point-mutation data obtained by using two unfolding methods each with 2,804 and 2,418 different point mutations, respectively, shows that most free energy changes are within ± ≈5.0 Kcal/mol.[3] The change in thermodynamic stability upon mutation can be computed as $\Delta\Delta G^D = \Delta G^D_m - \Delta G^D_{wt}$, where $\Delta G^D$ is the free energy of denaturation and *m* and *wt* stand for the *mutant* and *wild type* protein, respectively. Considering that mutations affect mostly the native state of proteins[3] the above-mentioned range of variation, $\Delta\Delta G^D < \pm \approx 5$ Kcal/mol, is consistent with the determined upper bound limit for the marginal stability of proteins (≈7.4 Kcal/mol). As a result, the very existence of a universal stability upper bound of the marginal stability provides a physical substrate for neutral evolution to occur. Indeed, if mutations introduce fluctuations greater than the stability upper bound then the bio-molecule (or bio-macromolecular complexes) will unfold. The remaining mutations could only marginally destabilize/stabilize a bio-molecule structure. Thus, from a thermodynamics perspective, most mutations and PTMs retaining biological function should be neutral[17] or nearly neutral.[18] Inclusion of PTMs in this analysis is a reasonable assumption because, from the stability point of view, the effect of PTMs should be similar to a point-mutation. If the stability upper bound is universal then the neutral evolution should also be (with respect to protein stability) a universal phenomenon; thus, not directly related to any details, such as the chemical composition, size or architecture, of the bio-molecule. As far as we know this particular connection between marginal stability and neutral evolution has been previously understudied.[3-12]

3. While we are not concerned in studying the protein free-energy of denaturation, but rather their marginal stability, it is worth mentioning that the proposed marginal stability upper bound of ≈7.4 Kcal/mol is lower, and thus consistent, with the observed average free-energy of denaturation ($\Delta G^D$) of nine monomeric globular proteins, i.e., ≈8 ≤ $\Delta G^D$ ≤ ≈14 Kcal/mol.[15]

So far, the aforementioned reasoning argues that the variation in the marginal stability for different proteins and bio-macromolecular complexes should be narrowly restricted to similar values but does not explain why the stability should be marginal. The nature of the latter can be understood intuitively after considering that the largest free-energy difference ($\Delta G$) among coexisting native-like states should verify two conditions:[13] first, $\Delta G > 1$ because the native state is the lowest free-energy conformation, i.e., the Anfinsen's thermodynamic hypothesis,[2] and second, $\Delta G$ should be small because "…*all conformers in equilibrium with the native-state will posses higher, although comparable, free energy…*"[13] This second condition, together with existing experimental evidence (see item 3 above), suggests that at the global free-energy minimum the stability of proteins is governed by fluctuations (jiggling and wiggling of atoms)[19] due to the interplay of pairwise and many-body interactions on both the proteins and the solvent and, therefore, should be marginal. At this point, is worth noting that intrinsically disordered proteins, prion proteins and metamorphic proteins present challenges to the Anfinsen's thermodynamic hypothesis. Therefore, caution must be exercised in the generalization of our results for such cases.

It is interesting to note that Taverna & Goldstein[5] propose a geometrical argument to rationalize the marginal stability of proteins. Indeed, in their work the authors propose that, because the protein sequence space is of high-dimensionality, stable proteins are restricted to live in manifold (or a marginally stable fringe region). How this geometrical explanation is related to our thermodynamic argument, or even if such connection exists, is an open and interesting question that may deserve further attention.

The apparent irrelevance on biological features, that follows from our statistical thermodynamics analysis, does not mean that important variables, such as the effective population size, have a null effect shaping the evolution of proteins or other biomolecules. It only means that the protein marginally stability upper bound has a physical origin. The behavior of the systems below that upper bound may indeed have multicausal origins. For example, Goldstein[12] discuss how protein stability depends on a mutation-selection balance that in turn depends on the effective population size. Combining their results with ours seems to imply the existence of an upper bound to the influence that the effective population size could have on the stability of proteins. Additionally, we want to highlight that, in line with Goldstein[11,12] and Taverna & Goldstein,[5] our results provide sound evidence that marginal stability is not an adaptive feature of proteins and other biomolecules.

Overall, here we argue that marginal stability of proteins is essentially a consequence of the Anfinsen's thermodynamics hypothesis and thus not a consequence of evolution, but rather the physical substrate for (neutral) evolution to occur. Additionally, the marginal stability also seems to be a universal property of bio-molecules and macro-molecular complexes. Taken all together these observations imply that a neutral or nearly neutral evolution should also be, with respect to the stability, a universal property of proteins and bio-molecular complexes.